\documentclass[aps,prl,twocolumn,superscriptaddress,nofootinbib,longbibliography]{revtex4-2}
\usepackage{amsmath,amssymb,bm,mathtools}
\usepackage{physics}
\usepackage{microtype}
\usepackage{hyperref}
\usepackage{xcolor}
\usepackage{booktabs,array}
\hypersetup{colorlinks=true,linkcolor=blue,citecolor=blue,urlcolor=blue}
\newcommand{\Ad}{\operatorname{Ad}}
\newcommand{\Lie}{\operatorname{Lie}}
\newcommand{\Alg}{\operatorname{Alg}}
\newcommand{\Mat}{\operatorname{Mat}}
\newcommand{\Sym}{\operatorname{Sym}}
\newcommand{\spec}{\operatorname{spec}}
\newcommand{\KL}{D_{\mathrm{KL}}}
\newcommand{\Fisher}{\mathcal I}
\newcommand{\E}{\mathbb E}
\newcommand{\Pp}{\mathbb P}
\begin{document}

\title{Chaos is Target-Blind in High-Energy QCD Evolution}
\author{Raktim Abir}
\email{raktim.ph@amu.ac.in}
\affiliation{Department of Physics, Aligarh Muslim University, Aligarh-202001, India}
\date{August 27, 2026}

\begin{abstract}
Chaotic evolution measures how rapidly nearby configurations separate, but the existence of such rapid evolution does not by itself determine what physical information the evolution actually carries. For fixed-coupling JIMWLK evolution with fresh transverse-local, color-isotropic noise, we show that this distinction becomes exact - leading to a no go theorem. At finite rapidity step and finite transverse regulator, the exact linearized update depends on the Wilson-line background only through local adjoint rotations of fresh color-isotropic noise. Once the past is conditioned upon, these rotations leave the complete local noise measure invariant, so the tangent process has the same probability law for any target ensemble evolved with the same regulated fixed-coupling dynamics, including dilute and saturated ensembles. After the exact global color-rotation mode is removed, the reduced tangent process found to have  a positive leading Lyapunov exponent which indicate chaotic evolution. Fixed-coupling JIMWLK therefor can be dynamically chaotic while its tangent-only statistics carry no information about the target ensemble. Spatially nonlocal noise however evades the argument because correlations between distinct sites leave relative Wilson lines in the conditional covariance.
\end{abstract}
\maketitle

 \textit{Introduction:--}
Chaotic dynamics is usually associated with extreme sensitivity to initial conditions: nearby trajectories separate rapidly, and Lyapunov exponents quantify the asymptotic rate of that separation~\cite{Ott}. But this rapid separation raises a different question as well: does stronger sensitivity to the initial condition also reveal more about the physical state through which the evolution proceeds? The two ideas are not equivalent. \textit{Sensitivity} describes how rapidly nearby evolutions separate, whereas \textit{identifiability} asks whether observing that separation allows us to learn which physical state is being evolved. High-energy QCD provides a setting in which these two notions can be separated exactly.

In a deep-inelastic scattering experiment at very high collision energy, partons carrying only a small fraction $x$ of the hadron's longitudinal momentum become increasingly important. In this small-$x$ regime the gluon density grows, nonlinear interactions among the gluons become unavoidable, and the target approaches the regime of gluon saturation. The transition between dilute and dense gluonic matter is characterized by the saturation scale $Q_s$. JIMWLK evolution describes how the Wilson-line ensemble representing the color field of the hadronic target changes as $x$ decreases, or equivalently as the rapidity $Y=\ln(1/x)$ increases~\cite{MV,Balitsky,Kovchegov,JIMWLK,IancuJIMWLK,IancuJIMWLKII,Weigert2002,IancuReview,WeigertReview,Rummukainen,IancuTriantafyllopoulos,EnbergMunier}. Ordinary JIMWLK observables retain information about whether the target is dilute or saturated through Wilson-line correlators and their associated transverse scales. This raises a different possibility: instead of reading the target directly from those background observables, could one identify it from the way an infinitesimal disturbance grows while the evolution proceeds?

One may therefore start with two nearly identical JIMWLK configurations and follow how their separation grows with rapidity. This poses two distinct questions. \textit{First, is the JIMWLK flow dynamically unstable, in the sense that infinitesimal perturbations grow exponentially? Second, do the statistics of that growth retain information about whether the underlying target is dilute or saturated?} The first question concerns \textit{chaos}; the second concerns \textit{inference}. They need not have the same answer: a random product of linear maps can generate strong stretching even when the probability law of that stretching is independent of the background one hopes to diagnose.

The central result of this Letter is precisely such a separation. For standard fixed-coupling JIMWLK evolution with fresh transverse-local, color-isotropic noise, we prove an {\it exact no-go theorem} for the second question: the complete tangent stochastic process has the same probability law for every Wilson-line target ensemble evolved with the same regulated fixed-coupling dynamics. A tangent-only observer therefore cannot determine whether the target is dilute or saturated, cannot infer $Q_s$, and cannot recover that information simply by waiting longer or by applying a more elaborate estimator. At the same time, the answer to the first question is yes. After the exact global color-rotation zero mode is removed, the reduced tangent dynamics is noncompact and has a positive leading Lyapunov exponent. Fixed-coupling JIMWLK is therefore chaotic while remaining target-blind at the level of tangent-only statistics. Here ``tangent-only'' means that the observer has access only to the evolution of the infinitesimal perturbation, without separately supplying the underlying Wilson-line configuration or another target-sensitive observable. 

The no-go statement concerns information generated by the perturbation dynamics itself. A norm, initialization rule, or postprocessing procedure that is supplied with $Q_s$, a Wilson-line correlator, or another target-sensitive quantity can of course discriminate between targets, but the information then enters through that additional input. Subject to this separation, the result is stronger than the failure of a particular Lyapunov estimator: every measurable functional of the tangent history has the same law for all target ensembles evolved under the stated fixed-coupling dynamics.

The proof proceeds in three steps. We first derive the exact finite-step evolution of an infinitesimal perturbation and show how the Wilson-line background enters its Jacobi map. We then use the local color isotropy of the fresh noise to prove that the complete tangent law is independent of the target ensemble, and separately establish that the reduced tangent dynamics is unstable. Finally, we show how spatially nonlocal noise evades the no-go mechanism by allowing relative Wilson lines to survive in the conditional covariance~\cite{LappiMantysaari,Altinoluk}.

\textit{Exact evolution of infinitesimal perturbations.--}
We keep both the rapidity step and the transverse lattice finite. The key probability identity can then be established directly for the regulated Langevin update, before taking any continuum limit. Let $a=\sqrt{\delta Y}$. A symmetric fixed-coupling JIMWLK step may be written as~\cite{Rummukainen,LappiMantysaari}
\begin{align}
U_x^+ &= e^{-iaA_x}U_xe^{iaB_x}, \label{eq:update}\\
A_x &= c\sum_{z,i}K^i_{x-z}X_z^i,
\qquad X_z^i=U_z\xi_z^iU_z^\dagger, \label{eq:Adef}
\end{align}
where $K^i_r=r^i/r^2$, color indices are suppressed, and $B_x$ is obtained from $A_x$ by replacing $X$ with the unrotated noise $\xi$. The fresh noise is independent from one rapidity step to the next and local in the transverse plane,
\begin{equation}
\E\,\xi_{z}^{i,a}\xi_{z'}^{j,b}=\delta^{ij}\delta^{ab}\delta_{zz'}. \label{eq:localnoise}
\end{equation}
The normalization $c$ contains the conventional fixed-coupling factors and is immaterial for the argument.

We now perturb the Wilson line from the left, $U_x'=e^{i\epsilon_x}U_x$, with $\epsilon_x$ infinitesimal in the Lie algebra. Since the Langevin step itself is finite, it is important to differentiate the matrix exponential before making any small-step expansion. Duhamel's identity then gives the exact one-step Jacobi map
\begin{align}
\epsilon_x^+&=\Ad_{e^{-iaA_x}}\epsilon_x
-a\int_0^1ds\,\Ad_{e^{-isaA_x}}\delta A_x, \label{eq:jacobi}\\
\delta A_x&=c\sum_{z,i}K^i_{x-z}\,i[\epsilon_z,X_z^i]. \label{eq:dA}
\end{align}
The right rotation in Eq.~\eqref{eq:update} is the same in the perturbed and unperturbed updates and cancels; the derivation is given in SM Sec.~\ref{Ssec:jacobi}. The remaining background dependence is very restricted: in the Jacobi map it appears only through the freshly rotated variables $X_z^i$. For a fixed realization of the raw noise, $X_z^i=U_z\xi_z^iU_z^\dagger$ still changes with the Wilson-line configuration, and so does $M(X)$. Equation~\eqref{eq:jacobi} therefore confines the background dependence without removing it. The conditional probability law of $X$ determines whether that remaining dependence survives statistically.

The physical mechanism can be made simpler in this form: the target acts on each fresh local noise variable only through an adjoint color rotation, while an isotropic Gaussian law is invariant under that rotation. We now formulate this statement conditionally, which is necessary because the rotation itself is determined by the evolved Wilson lines. Let $\mathcal F_Y$ denote the complete history before the current rapidity step. The Wilson lines $U(Y)$ are then $\mathcal F_Y$-measurable, whereas the fresh noise $\xi(Y)$ is independent of $\mathcal F_Y$. In adjoint components,
\begin{equation}
X_z^{i,a}=R_z^{ab}(U)\,\xi_z^{i,b},\qquad R_z=\Ad_{U_z}. \label{eq:adjrot}
\end{equation}
For a compact gauge group, the adjoint action is orthogonal with respect to the invariant Lie-algebra metric. Because the covariance in Eq.~\eqref{eq:localnoise} is also local, this orthogonal rotation may be carried out independently at every site and polarization. Conditioned on the past, the rotated noise therefore has the same measure as the original noise,
\begin{equation}
\Pp[X(Y)\mid\mathcal F_Y]=\Pp[\xi(Y)]. \label{eq:conditional}
\end{equation}
Equation~\eqref{eq:conditional} gives equality of the complete conditional Gaussian measures. Once the past is fixed, the matrices $R_z(U)$ form a block-diagonal orthogonal transformation acting independently at each site and polarization. The Gaussian quadratic form and the measure are unchanged. Equivalent measure-level checks are given in SM Sec.~\ref{Ssec:noise}.

\textit{Target-blindness theorem.--}
For a fixed realization of $X$, Eq.~\eqref{eq:jacobi} is linear in $\epsilon$. Let $M(X)$ denote the corresponding tangent matrix on the quotient space obtained after removing the uniform symmetry mode. Since the conditional law of $X$ is independent of the background, the same is true of the one-step tangent kernel:
\begin{equation}
\Pp(\epsilon^+\mid\epsilon,U)=\Pp(\epsilon^+\mid\epsilon). \label{eq:kernelblind}
\end{equation}
The one-step tangent transition kernel is independent of the Wilson lines present at that step.

Because the same conditional statement holds at every fresh-noise step, it can be iterated. At step $n$, the entire previous evolution fixes the current Wilson lines and tangent propagator, but the next innovation is again fresh. Equation~\eqref{eq:conditional} therefore supplies the same conditional distribution for the next tangent matrix regardless of which target ensemble produced that past. The background may affect the realized Wilson-line trajectory, but it never changes the conditional tangent transition kernel. Iterating the same conditional kernel makes every finite cylinder probability of the tangent process independent of the target ensemble. If
\begin{equation}
J_0=\mathbf 1,\qquad J_{n+1}=M(X_n)J_n, \label{eq:propagator}
\end{equation}
then for any two initial target ensembles $W_1[U]$ and $W_2[U]$ evolved with the same regulated fixed-coupling dynamics,
\begin{equation}
\Pp_{W_1}(J_0,\ldots,J_N)=\Pp_{W_2}(J_0,\ldots,J_N). \label{eq:pathlaw}
\end{equation}
The equality holds at finite lattice spacing and finite rapidity step, with fixed ultraviolet and infrared regulators. If a continuum limit exists, equality of the finite-dimensional tangent distributions passes to that limit; the corresponding weak-limit and It\^o formulations are described in SM Sec.~\ref{Ssec:finite-continuum}.

Equation~\eqref{eq:pathlaw} is the central result. The full Jacobi history has the same probability law for the two target ensembles. Any quantity constructed from this history alone inherits the same independence. This includes finite-time singular values, Lyapunov estimators, stopping-time constructions, large-deviation observables, and the asymptotic Lyapunov spectrum. In particular,
\begin{equation}
\spec_{\rm Lyap}[W_1]=\spec_{\rm Lyap}[W_2]. \label{eq:lyapspec}
\end{equation}
Thus, when the regulated fixed-coupling evolution law is held fixed, replacing a dilute target ensemble by a saturated one does not change any distribution obtained solely from the perturbation history.

The same statement has an immediate information-theoretic form. Let $\Theta$ label a family of target ensembles and let $T_N$ denote any data obtained from $J_{0:N}$ using a metric and postprocessing rule that do not themselves contain background information. For any two values $\theta_1$ and $\theta_2$,
\begin{equation}
\KL[P(T_N|\theta_1)\Vert P(T_N|\theta_2)]=0. \label{eq:info}
\end{equation}
The two tangent-data distributions are identical. Mutual information and Fisher information vanish as consequence (SM Sec.~\ref{Ssec:info}). A nonlinear estimator, including a machine-learning map, can reorganize the tangent data but cannot create target information that is absent from its probability law. Increasing the observation interval does not change this conclusion.

The restriction to tangent-only data is essential to the physical statement. If the initialization, norm, metric, or postprocessing is supplied with $U$, $Q_s$, a dipole amplitude, or another target-sensitive quantity, the resulting diagnostic may of course distinguish target ensembles. In that case, however, the information has entered through the additional input. The theorem concerns only information generated by the tangent dynamics itself.

 Two backgrounds driven by the same stored realization of the raw noise $\xi$ generally produce different tangent trajectories because they rotate that realization differently. Equation~\eqref{eq:pathlaw}, by contrast, compares the marginal tangent laws, which are identical.

\textit{Chaotic growth without target information.--}
Target-blindness would have been trivial if the tangent dynamics were only neutral or compact. We therefore examine the growth only after the exact finite-step probability statement has been established. Expanding the Jacobi map to leading order in $\sqrt{\delta Y}$ gives the tangent drivers
\begin{equation}
(B_{zia}\epsilon)_x=icK^i_{x-z}[t^a,\epsilon_z-\epsilon_x]. \label{eq:driver}
\end{equation}
A spatially uniform tangent is annihilated by Eq.~\eqref{eq:driver}. It is the infinitesimal form of a global color rotation of all Wilson lines and is therefore an exact zero mode of the regulated evolution. The stability question has to be ideally posed on the quotient by this symmetry direction.

On this quotient space, the relevant question is whether the random linear maps remain inside a compact rotation group or generate noncompact directions. A product of random rotations can be complicated without producing exponential stretching; positive Lyapunov growth requires directions that can change tangent-vector lengths. Here those directions are generated by the interplay of the spatial kernel and the non-Abelian color algebra. For lattices with at least three distinct sites, the spatial generators together with the color commutator structure generate the full special-linear tangent algebra for SU(2) and SU(3). The spatial closure, the SU(2) and SU(3) Jordan closures, and their combination are given in SM Secs.~\ref{Ssec:spatial}--\ref{Ssec:fullclosure}. The support of the reduced random evolution is consequently not confined to a compact subgroup: it contains shear and stretching directions on the quotient tangent space. Under the corresponding irreducibility and support conditions established there, standard results for noncommuting random matrix products~\cite{Furstenberg} give
\begin{equation}
\lambda_1>0. \label{eq:positive}
\end{equation}
where $\lambda_1$ is the largest (leading) Lyapunov exponent of the reduced tangent dynamics, after removing the exact global color-rotation zero mode.
The reduced tangent process therefore has exponential growth, while the probability law governing that growth remains independent of the target ensemble.

The Lyapunov spectrum satisfies a complementary constraint. In the continuum Stratonovich formulation, the JIMWLK emission fields are sums of left- and right-invariant vector fields on the product of site groups. Their divergence with respect to product Haar measure vanishes. The stochastic flow therefore preserves this volume pathwise, implying
\begin{equation}
\sum_n\lambda_n=0. \label{eq:sumszero}
\end{equation}
A positive exponent is accompanied by contraction in other tangent directions. The base trajectory remains on the compact Wilson-line group manifold, while its derivative acts in a linear tangent space whose cocycle can be noncompact.

\textit{Scope of the statement.--}
The theorem compares different target ensembles under the same stochastic evolution. Its assumptions are therefore part of the physical content of the result. The noise must be fresh in rapidity, local in transverse position, and invariant under local adjoint color rotations; the background dependence of the Jacobi map must enter through those freshly rotated increments. The diagnostic must also be tangent-only in the sense defined above. 

Target dependence must also be separated from dependence on the evolution law. The Lyapunov spectrum can change if one changes $\alpha_s$, the kernel, lattice geometry, ultraviolet or infrared regulator, or rapidity convention. Such changes modify the stochastic dynamics itself. They do not show that, for a fixed dynamics, the tangent process can discriminate between two Wilson-line ensembles evolved by that same dynamics.

The same separation fixes the scope of possible counterexamples. A background-dependent norm may vary strongly across the saturation region, but then the background has entered the measurement directly. An estimator that is trained on both tangent data and a dipole observable may classify targets, but the classification need not originate from the tangent instability. The theorem asks the narrower question in which the perturbation history is the only information channel. Within that channel, Eq.~\eqref{eq:info} gives zero target information.

\textit{How blindness is broken.--}
The proof also shows where target sensitivity can first return. The key step in Eq.~\eqref{eq:conditional} used transverse locality as well as color isotropy. Suppose instead that the raw noise has a nonlocal spatial covariance,
\begin{equation}
\E\,\xi_z^{i,a}\xi_{z'}^{j,b}=\delta^{ij}\delta^{ab}\alpha(z-z'), \label{eq:nonlocal}
\end{equation}
so that noise at two distinct sites is correlated. The two local adjoint rotations can no longer be absorbed independently. Conditioned on the Wilson lines,
\begin{equation}
\E[X_z^{i,a}X_{z'}^{j,b}|U]
=\delta^{ij}\alpha(z-z')\,\Ad^{ab}_{U_zU_{z'}^\dagger}. \label{eq:relativelink}
\end{equation}
A relative adjoint Wilson line now survives in the tangent covariance. At equal sites, $z=z'$, the two rotations combine to the identity and the local cancellation is recovered. At unequal sites, however, the covariance ties together two independently rotated noises, and the two rotations combine into $\Ad_{U_zU_{z'}^\dagger}$ rather than disappearing. The object that remains is therefore a bilocal property of the target itself. This is the first background-sensitive quantity that escapes the local-noise proof.

Equation~\eqref{eq:relativelink} leaves a possible carrier of target information. Running-coupling Langevin formulations generate spatially nonlocal correlations of this kind~\cite{LappiMantysaari,Altinoluk} and therefore lie outside the fixed-coupling theorem. The presence of the relative link alone does not imply a universal ``Lyapunov saturation scale.'' The induced response may remain regulator dependent or may fail to organize around a unique momentum proportional to $Q_s$. Establishing such a positive construction requires a specified nonlocal covariance and an observable whose scaling can be demonstrated independently of the regulator.

\textit{Discussion.--}

The possibility has a direct precedent in the small-$x$ literature. Earlier. Kharzeev and Tuchin found that a discrete version of the mean-field Kovchegov evolution can become chaotic when the linear small-$x$ growth is sufficiently rapid, and Tuchin subsequently emphasized that the chaotic regime appears in a phenomenologically interesting region~\cite{KharzeevTuchin,TuchinChaos,ZhuShenRuan2008,ZhuShenRuan2016}. Their setting is a discretized mean-field BK evolution rather than the stochastic Wilson-line dynamics considered here. The earlier results nevertheless make the connection between chaos and saturation physics concrete and motivate asking whether the instability of JIMWLK itself can serve as a dynamical diagnostic of the target.

The mechanism behind the statement presented here is rather simple in physical terms: the target rotates each fresh local color fluctuation, but an isotropic noise distribution cannot tell that such a rotation has occurred.

The no-go statement applies beyond any particular definition of a Lyapunov exponent. At fixed coupling, with fresh transverse-local color-isotropic noise, the complete tangent-history law is independent of the target ensemble. Lyapunov exponents, finite-rapidity growth distributions, large deviations, stopping rules, and other tangent-only estimators are different functionals of this same law and inherit the same blindness. If two target ensembles evolved with the same regulated dynamics give different tangent-growth distributions, then either an assumption of the theorem has been left or target-sensitive information has entered through the diagnostic itself.

The tangent dynamics nevertheless remains nontrivial: on the reduced space it has a positive leading Lyapunov exponent. The result separates exponential sensitivity from target information within the same evolution. The failure of the proof identifies the required structural change. A chaos-based saturation diagnostic must involve some structure that is not removed by independent local adjoint rotations of the fresh noise. Spatially nonlocal covariance is one such possibility because it retains relative Wilson lines. A chaos-based diagnostic must therefore test whether the probability law governing the growth retains a background-sensitive object or not.

\bibliographystyle{apsrev4-2}
\bibliography{ref.bib}

\clearpage
\onecolumngrid
\setcounter{equation}{0}
\renewcommand{\theequation}{S\arabic{equation}}
\renewcommand{\theHequation}{S\arabic{equation}}
\newcounter{smsection}
\renewcommand{\thesmsection}{\Roman{smsection}}
\newcounter{smsubsection}[smsection]
\renewcommand{\thesmsubsection}{\Roman{smsection}.\Alph{smsubsection}}
\renewcommand{\theHsmsubsection}{\arabic{smsection}.\arabic{smsubsection}}
\newcommand{\SMsection}[2]{%
  \refstepcounter{smsection}%
  \section*{\Roman{smsection}.\ #1}%
  \label{#2}%
}
\newcommand{\SMsubsection}[2]{%
  \refstepcounter{smsubsection}%
  \subsection*{\Alph{smsubsection}.\ #1}%
  \label{#2}%
}

\begin{center}
{\large\bf Supplemental Material for ``Chaos is Target-Blind in High-Energy QCD Evolution''}\\[8pt]
Raktim Abir\\[2pt]
{\it Department of Physics, Aligarh Muslim University, Aligarh, India}
\end{center}
\vspace{0.5em}
\SMsection{Physical setting and statement of the result}{Ssec:physical}
At high collision energy, the small-$x$ gluon density of a hadron or nucleus grows until nonlinear interactions become important.  The scale that separates the dilute and dense regimes is the saturation scale $Q_s(Y)$, with $Y=\ln(1/x)$.  In the Color Glass Condensate description, the target is represented by an ensemble of Wilson lines $U_x\in SU(N_c)$, one at each transverse position.  Correlators of these Wilson lines contain the physical information that distinguishes one target ensemble from another.

JIMWLK evolution specifies how that ensemble changes with rapidity.  Earlier studies of a discrete mean-field BK/Kovchegov evolution found chaotic behavior when the linear small-$x$ growth is sufficiently rapid and in a phenomenologically interesting kinematic region.  That setting differs from the stochastic Wilson-line evolution considered here, but it motivates a sharper question for JIMWLK.  If two infinitesimally close Wilson-line configurations are evolved with the same realization of the noise, does their separation grow exponentially, and can the statistics of that growth reveal whether the underlying target is dilute or saturated?

The two questions lead to different calculations.  The target-blindness statement follows from the exact finite-step evolution of an infinitesimal perturbation.  Once the common right rotation has canceled, the Wilson-line background enters the Jacobi map only through the rotated fresh noise $X_z^i=U_z\xi_z^iU_z^\dagger$.  Conditioning on the previous rapidity history fixes the Wilson lines, while the next noise increment remains fresh and color isotropic.  The adjoint rotation then leaves its probability law unchanged.  Iterating this conditional statement gives a tangent path measure that is independent of the target ensemble.

The instability question requires a different argument.  A common color rotation of all sites is an exact neutral direction and is removed first.  On the remaining tangent space, the spatial kernel matrices and the adjoint color generators generate the full special-linear algebra under the nondegeneracy conditions stated below.  The resulting random matrix process is noncompact and irreducible, which gives a positive leading Lyapunov exponent.  The continuum flow also preserves Haar volume, so the positive expansion is balanced by contraction in other directions.

Together these results show that fixed-coupling JIMWLK can possess Lyapunov instability even though tangent-only data carry no information about which target ensemble is being evolved.  The derivation is carried out first for a finite transverse regulator and finite rapidity step.  The continuum formulation is considered only after the regulated probability statement has been established.

\SMsubsection{Assumptions}{Ssec:assumptions}

We assume fresh rapidity noise, transverse locality of its fixed-coupling covariance, and invariance of the color distribution under the adjoint action at each site.  The derivation below shows that the Wilson-line background enters the exact Jacobi map only through the rotated variables $X_z^i=U_z\xi_z^iU_z^\dagger$.  The tangent initialization, norm, normalization procedure, stopping rule, and postprocessing are chosen independently of the target ensemble.

A statistic is ``tangent-only'' when all of its inputs are constructed from the tangent propagator history together with fixed external choices that do not encode the target.  Supplying the estimator with $U$, a dipole amplitude, a previously measured $Q_s$, or a norm built from target correlators introduces a separate information channel and therefore lies outside the theorem.

The theorem also keeps the regulated dynamics fixed when target ensembles are compared.  The Lyapunov exponents may depend on the coupling normalization, lattice geometry, ultraviolet and infrared regulators, the rapidity step, or the chosen background-independent tangent norm.  Those are properties of the evolution law or regulator.  Target sensitivity would require a change in the tangent probability law when the initial Wilson-line probability distribution is changed.

The conclusion changes when one of these ingredients is changed.  Spatially nonlocal noise correlates distinct transverse sites, so the background rotations can no longer be absorbed independently.  Color-anisotropic noise introduces a preferred color tensor, while rapidity-correlated noise introduces memory.  A tangent equation containing Wilson lines outside the rotated fresh-noise combination is likewise outside the argument.  The nonlocal case will be examined explicitly near the end of the Supplement.

\SMsection{Color algebra and adjoint rotations}{Ssec:color}
We use Hermitian generators $t^a$ of $SU(N_c)$ in the fundamental representation,
\begin{equation}
\operatorname{tr}(t^at^b)=\frac12\delta^{ab},
\qquad [t^a,t^b]=if^{abc}t^c.
\end{equation}
A Hermitian Lie-algebra element is written $A=A^at^a$, with real components $A^a$.  The natural invariant inner product is
\begin{equation}
\langle A,B\rangle=2\operatorname{tr}(AB)=A^aB^a.
\end{equation}
Thus the adjoint color space is an ordinary real Euclidean vector space of dimension $d_A=N_c^2-1$.

For $U\in SU(N_c)$, conjugation acts on the generators as
\begin{equation}
Ut^bU^\dagger=t^a(\Ad_U)_{ab}.
\end{equation}
Trace invariance gives
\begin{align}
\delta^{ab}
&=2\operatorname{tr}(t^at^b)\\
&=2\operatorname{tr}(Ut^aU^\dagger Ut^bU^\dagger)\\
&=(\Ad_U)_{ca}(\Ad_U)_{cb}.
\end{align}
Hence
\begin{equation}
(\Ad_U)^T\Ad_U=\mathbf 1.
\end{equation}
Because $SU(N_c)$ is connected and $\Ad_{\mathbf1}=\mathbf1$, the determinant cannot jump from $+1$ to $-1$; therefore $\Ad_U\in SO(d_A)$.

This group-theory property is what removes the Wilson-line dependence from the fresh-noise law.  At a fixed transverse site, conjugation by $U_z$ changes a particular realization of the color noise, but it only rotates that realization in the real adjoint space.  For an isotropic measure the rotated and unrotated variables have the same distribution once the Wilson line is fixed.  Thus $X_z^i$ and $\xi_z^i$ differ configuration by configuration while sharing the same conditional law.

  For $U(\theta)=e^{i\theta t^a}$,
\begin{equation}
U(\theta)t^bU^\dagger(\theta)
=t^b-\theta f^{abc}t^c+O(\theta^2),
\end{equation}
so the real adjoint generators may be written
\begin{equation}
(A^a)_{bc}=-f^{abc},
\qquad (A^a)^T=-A^a.
\end{equation}
They generate ordinary rotations of the real adjoint color space.

\SMsection{Regulated state space and JIMWLK update}{Ssec:lattice}
We regulate the transverse plane by a finite set of $L$ distinct sites.  A Wilson-line configuration is therefore a point of the compact product manifold
\begin{equation}
\mathcal M=SU(N_c)^L.
\end{equation}
Working at finite $L$ keeps every tangent space finite-dimensional and makes one rapidity step an ordinary smooth random map.  The central blindness result first established at this level before continuum limit is taken.

To compare nearby configurations we use a left-trivialized tangent coordinate,
\begin{equation}
U_x'=e^{i\epsilon_x}U_x,
\qquad \epsilon_x=\epsilon_x^at^a.
\end{equation}
To first order, $\delta U_x=i\epsilon_xU_x$ and therefore $\epsilon_x=-i\,\delta U_xU_x^\dagger$.  The product bi-invariant metric gives the background-independent norm
\begin{equation}
\|\epsilon\|^2=\sum_x2\operatorname{tr}(\epsilon_x^2)
=\sum_{x,a}(\epsilon_x^a)^2.
\end{equation}
The unreduced real tangent dimension is $Ld_A$.

There is an exact symmetry direction.  If $\epsilon_x=\epsilon_0$ at every site, the perturbed configuration differs from the original one by a common left color rotation, leaving the relative color structure across the transverse plane unchanged.  We therefore work on the quotient by this $d_A$-dimensional orbit.  A convenient representative satisfies
\begin{equation}
\sum_x\epsilon_x=0,
\end{equation}
obtained by subtracting the site average.  The reduced tangent dimension is
\begin{equation}
D=(L-1)(N_c^2-1).
\end{equation}
Later we verify directly from the exact finite-step map that this uniform direction has multiplier one, so the quotient is not based on a small-step approximation.

\SMsubsection{The regulated Weizs\"acker--Williams kernel}{Ssec:kernel}
For a nonzero two-dimensional displacement $r$, we use
\begin{equation}
K_r^i=\frac{r^i}{r^2},
\qquad K_0^i=0.
\end{equation}
The value at the coincident point is part of the finite regulator.  The kernel is odd, $K_{-r}^i=-K_r^i$, and is evaluated only on the finite set of lattice differences.  No rotational symmetry of the lattice is needed for the probability argument.

A later spatial-generation argument uses the fact that the map $r\mapsto r/r^2$ is injective away from the origin.  Indeed, if $k=r/r^2$, then $k^2=1/r^2$ and $r=k/k^2$.  Distinct nonzero displacements remain distinguishable through their kernel vectors.  This simple geometric fact is what allows the spatial algebra to retain information about which lattice sites are connected even on a generic finite lattice.

We use the following notation throughout.  $U_x$ denotes a Wilson line, $\xi_z^i$ the fresh local Lie-algebra noise, and $X_z^i=U_z\xi_z^iU_z^\dagger$ its adjoint rotation by the incoming background.  $A_x$ and $B_x$ are the left and right Lie-algebra fields in one symmetric Langevin step.  $\epsilon_x$ is a left-trivialized tangent perturbation.  $M(X)$ is the one-step Jacobi matrix on the reduced tangent space, and $J_n$ is the product of those matrices up to step $n$.
\SMsubsection{Regulated JIMWLK update}{Ssec:update}
Consider a finite transverse lattice with sites labelled by $x,z,\ldots$.  The standard symmetric fixed-coupling update can be written
\begin{equation}
U_x^+=L_xU_xR_x,
\qquad
L_x=e^{-iaA_x},\qquad
R_x=e^{iaB_x},
\label{S:update}
\end{equation}
where $a=\sqrt{\delta Y}$ and
\begin{align}
A_x&=c\sum_{z,i}K^i_{x-z}X_z^i,
&X_z^i&=U_z\xi_z^iU_z^\dagger,\label{S:A}\\
B_x&=c\sum_{z,i}K^i_{x-z}\xi_z^i.
\label{S:B}
\end{align}
The kernel is $K_r^i=r^i/r^2$ away from the coincident point, with the regulated convention $K_0=0$.  The index $i$ is the transverse polarization label.  We take the Lie-algebra generators Hermitian, $[t^a,t^b]=if^{abc}t^c$, and expand $\xi_z^i=\xi_z^{i,a}t^a$.

The fresh Gaussian noise is centered and independent from rapidity step to rapidity step.  At fixed coupling its transverse and color covariance is ultralocal,
\begin{equation}
\E\,\xi_z^{i,a}\xi_{z'}^{j,b}=\delta^{ij}\delta^{ab}\delta_{zz'}.
\label{S:covlocal}
\end{equation}
All dependence on the Wilson-line background in the left rotation $L_x$ therefore appears through the adjointly rotated noise $X_z^i$, whereas the right rotation is built from the raw noise.  This left-right asymmetry is the structural feature exploited by the proof.

The no-go result is established at finite lattice spacing and finite rapidity step.  Its probability identity is therefore a statement about the regulated update itself and does not rely on an It\^o/Stratonovich conversion or a small-step truncation.

\SMsection{Exact Jacobi map from Duhamel differentiation}{Ssec:jacobi}
Perturb one configuration from the left,
\begin{equation}
U_x'=e^{i\epsilon_x}U_x,
\qquad \epsilon_x=\epsilon_x^at^a,
\end{equation}
and retain only terms linear in the infinitesimal tangent $\epsilon$.  The rotated noise varies as
\begin{align}
X_z^{\prime i}
&=e^{i\epsilon_z}U_z\xi_z^iU_z^\dagger e^{-i\epsilon_z}\nonumber\\
&=X_z^i+i[\epsilon_z,X_z^i]+O(\epsilon^2),
\end{align}
so that
\begin{equation}
\delta A_x=c\sum_{z,i}K^i_{x-z}\,i[\epsilon_z,X_z^i].
\label{S:dA}
\end{equation}
The variation of the left exponential must be handled without expanding in $a$.  For a matrix $A$ and variation $\delta A$,
\begin{equation}
\delta e^{-iaA}
=-ia\int_0^1ds\,e^{-i(1-s)aA}(\delta A)e^{-isaA}.
\label{S:Duhamel}
\end{equation}
This follows, for example, by differentiating $e^{-ia(A+\tau\delta A)}$ with respect to $\tau$ at $\tau=0$ and inserting the standard integral representation for the derivative of an operator exponential.

The perturbed updated line is
\begin{equation}
U_x^{\prime +}=(L_x+\delta L_x)e^{i\epsilon_x}U_xR_x.
\end{equation}
Define the new tangent by $U_x^{\prime +}=e^{i\epsilon_x^+}U_x^+$.  Multiplication on the right by $(U_x^+)^{-1}=R_x^{-1}U_x^{-1}L_x^{-1}$ cancels the common right rotation.  To linear order in $\epsilon$,
\begin{equation}
i\epsilon_x^+=\delta L_xL_x^{-1}+L_x(i\epsilon_x)L_x^{-1}.
\end{equation}
Using Eq.~\eqref{S:Duhamel},
\begin{align}
\delta L_xL_x^{-1}
&=-ia\int_0^1ds\,e^{-i(1-s)aA_x}(\delta A_x)e^{-isaA_x}e^{iaA_x}\nonumber\\
&=-ia\int_0^1ds\,e^{-i(1-s)aA_x}(\delta A_x)e^{i(1-s)aA_x}.
\end{align}
Relabelling $1-s\to s$ gives the exact regulated Jacobi map
\begin{equation}
\epsilon_x^+=\Ad_{e^{-iaA_x}}\epsilon_x
-a\int_0^1ds\,\Ad_{e^{-isaA_x}}\delta A_x
\label{S:exactJacobi}
\end{equation}
with $\delta A_x$ from Eq.~\eqref{S:dA}.  There has been no expansion in $a=\sqrt{\delta Y}$.  The only approximation is the defining linearization in the tangent itself.

After the right rotation has canceled, every dependence on the nonlinear Wilson-line background in Eq.~\eqref{S:exactJacobi} occurs through $A_x$, $\delta A_x$, and hence through the rotated noises $X_z^i=U_z\xi_z^iU_z^\dagger$.  No other Wilson-line coefficient remains in the Jacobi map.

For a fixed realization, the map remains background dependent because $X_z^i$ still contains $U_z$.  Equation~\eqref{S:exactJacobi} has only confined that dependence to the collection $X$.  Whether it survives in the tangent statistics is decided by the conditional law of these rotated variables.

\SMsection{Conditional invariance of the fresh noise}{Ssec:noise}
Let $\mathcal F_Y$ denote the sigma-algebra generated by the complete history up to, but not including, the fresh noise at rapidity $Y$.  The Wilson lines $U_z(Y)$ are $\mathcal F_Y$-measurable.  The new $\xi_z^i(Y)$ are independent of $\mathcal F_Y$.  In adjoint components,
\begin{equation}
X_z^{i,a}=R_z^{ab}(U)\xi_z^{i,b},
\qquad R_z=\Ad_{U_z}.
\label{S:R}
\end{equation}
For a compact gauge group the invariant Lie-algebra inner product can be chosen positive definite.  The adjoint action preserves it, hence
\begin{equation}
R_z^TR_z=\mathbf 1.
\end{equation}
Conditioning on $\mathcal F_Y$ freezes every $R_z$ into a deterministic orthogonal matrix.

The equality of the conditional laws can be seen directly from the product Gaussian density.  Suppressing polarization labels,
\begin{equation}
p(\xi)\,d\xi\propto
\exp\left[-\frac12\sum_z\xi_z^T\xi_z\right]\prod_zd\xi_z.
\end{equation}
Under $X_z=R_z\xi_z$,
\begin{equation}
\sum_zX_z^TX_z=\sum_z\xi_z^TR_z^TR_z\xi_z=\sum_z\xi_z^T\xi_z,
\end{equation}
and
\begin{equation}
\left|\det\frac{\partial X}{\partial\xi}\right|
=\prod_z|\det R_z|=1.
\end{equation}
Thus both the Gaussian weight and the integration measure are unchanged.

The same conditional statement follows from the characteristic functional.  For arbitrary test vectors $j_z^i$,
\begin{align}
\E\left[e^{i\sum_{zi}(j_z^i,X_z^i)}\big|\mathcal F_Y\right]
&=\E\left[e^{i\sum_{zi}(R_z^Tj_z^i,\xi_z^i)}\right]\nonumber\\
&=\exp\left[-\frac12\sum_{zi}\|R_z^Tj_z^i\|^2\right]\nonumber\\
&=\exp\left[-\frac12\sum_{zi}\|j_z^i\|^2\right].
\end{align}
This is the characteristic functional of the unrotated noise.  Therefore
\begin{equation}
\Pp[X(Y)\mid\mathcal F_Y]=\Pp[\xi(Y)].
\label{S:condlaw}
\end{equation}
Equation~\eqref{S:condlaw} is an equality of complete conditional measures, not only of their first two moments.  Whatever nonlinear history produced the present Wilson lines, the next rotated increment has the same Gaussian law as a fresh unrotated increment.

Conditioning separates the already-evolved background from the next stochastic increment.  In the unconditional joint distribution, $X$ and $U$ share the same history and need not be independent.  Once the past is fixed, $U$ is fixed as well, while the new increment remains a fresh isotropic Gaussian.  The Wilson line then acts only as an orthogonal matrix on that increment.  The resulting statement is therefore an identity for the conditional innovation law.

Gaussianity is convenient but not fundamental.  Any fresh local noise measure invariant under the adjoint action gives the same conclusion.  Locality permits an independent orthogonal transformation at every transverse site, and color isotropy makes those transformations measure preserving.

\SMsubsection{Locality and color isotropy}{Ssec:sharp}
For centered Gaussian noise let $C_{zz'}$ be the color covariance matrix, with polarization labels left implicit.  Invariance under arbitrary independent adjoint rotations would require
\begin{equation}
R_zC_{zz'}R_{z'}^T=C_{zz'}
\label{S:Cinv}
\end{equation}
for all choices of $R_z$ and $R_{z'}$ in the adjoint representation.  If $z\neq z'$, set $R_{z'}=\mathbf 1$ and vary $R_z$.  The columns of $C_{zz'}$ would then have to be invariant vectors of the adjoint representation.  A compact simple Lie algebra has no nonzero invariant vector in its adjoint representation, so
\begin{equation}
C_{zz'}=0,\qquad z\neq z'.
\end{equation}
At a single site, Schur's lemma implies that an invariant covariance block is proportional to the invariant metric.  Apart from polarization mixing and independent simple factors, a centered Gaussian covariance invariant under arbitrary sitewise adjoint rotations is therefore local in the transverse labels and isotropic in color.  These are not merely convenient choices; they are the covariance conditions that make the sitewise rotation invisible.

\SMsection{Target-blind tangent process}{Ssec:markov}
For a fixed noise realization $X$, Eq.~\eqref{S:exactJacobi} is a linear map in $\epsilon$.  Let $M(X)$ denote the corresponding matrix after quotienting the exact uniform zero mode.  Because $M$ depends on the nonlinear background only through $X$, Eq.~\eqref{S:condlaw} gives
\begin{equation}
\Pp(\epsilon^+\in d\eta\mid\epsilon,U)
=\Pp(M(X)\epsilon\in d\eta\mid\epsilon),
\end{equation}
with no residual dependence on $U$.  Equivalently,
\begin{equation}
\Pp(\epsilon^+\mid\epsilon,U)=\Pp(\epsilon^+\mid\epsilon).
\label{S:Markovblind}
\end{equation}
Equation~\eqref{S:Markovblind} gives the background-independent transition law for a single rapidity step.  Because the next noise increment is fresh after every possible previous history, the same transition law applies at each subsequent step.  No additional approximation is introduced by this iteration.

\SMsubsection{Complete propagator path law}{Ssec:pathlaw}
Introduce the full Jacobi propagator rather than following only one chosen tangent vector,
\begin{equation}
J_0=\mathbf 1,
\qquad J_{n+1}=M(X_n)J_n.
\label{S:Jrec}
\end{equation}
Let $K(J,dJ')$ be the corresponding one-step Markov kernel.  Equation~\eqref{S:condlaw} implies that $K$ is the same for every target ensemble.  For a cylinder set $A_0\times\cdots\times A_N$,
\begin{align}
\Pp_W(J_0\in A_0,\ldots,J_N\in A_N)
&=\int_{A_0}\delta_{\mathbf1}(dJ_0)
\prod_{n=0}^{N-1}K(J_n,dJ_{n+1})
\prod_{r=1}^N\mathbf1_{A_r}(J_r).
\label{S:cylinder}
\end{align}
No factor on the right contains the Wilson-line distribution $W[U]$.  Therefore, for any two initial target ensembles,
\begin{equation}
\Pp_{W_1}(J_0,\ldots,J_N)=\Pp_{W_2}(J_0,\ldots,J_N).
\label{S:pathlaw}
\end{equation}
The same path law therefore governs any tangent-only statistic constructed from the propagator with a fixed, background-independent prescription.

Let $F$ be any measurable map of the propagator path and define $T_N=F(J_0,\ldots,J_N)$.  The map may implement QR decomposition, Gram--Schmidt reorthogonalization, norm increments, singular-value decomposition, projective directions, finite-time exponent histograms, stopping rules, or randomized postprocessing independent of the target.  Equality of measures is preserved under pushforward, so
\begin{equation}
\Pp(T_N\in A\mid\theta_1)=\Pp(T_N\in A\mid\theta_2)
\end{equation}
for every measurable set $A$ and every pair of target parameters $\theta_1,\theta_2$.

The path theorem is also a statement about quenched Lyapunov exponents.  A quenched exponent is computed from a long product of random Jacobi matrices along a single realization, not from an annealed average of those matrices.  Equation~\eqref{S:pathlaw} does not replace the product by an average.  It says that the probability measure from which the entire random product is drawn is identical for all target ensembles.  Hence its almost-sure exponent, finite-time fluctuations, and rare-event tails are all background independent.

\SMsubsection{Information-theoretic corollaries}{Ssec:info}
Suppose the family of targets is labelled by a parameter $\Theta$.  This parameter may encode an initial saturation scale, a color-charge density, a nuclear size, or more generally a family of Wilson-line probability distributions.  Let $T_N$ be any tangent-only data produced from $J_{0:N}$ with a fixed background-independent metric.  From Eq.~\eqref{S:pathlaw},
\begin{equation}
p(t\mid\theta_1)=p(t\mid\theta_2)
\end{equation}
for all $t$ and all values of the target parameter.  The Kullback--Leibler divergence therefore vanishes identically,
\begin{equation}
\KL[p_{\theta_1}\Vert p_{\theta_2}]
=\int dt\,p(t\mid\theta_1)\ln\frac{p(t\mid\theta_1)}{p(t\mid\theta_2)}=0.
\end{equation}
If $\Theta$ is itself random, then $T_N$ is independent of it and
\begin{equation}
I(\Theta:T_N)=0.
\end{equation}
For a differentiable family,
\begin{equation}
\Fisher_\theta[T_N]
=\int dt\,p(t\mid\theta)\left[\partial_\theta\ln p(t\mid\theta)\right]^2=0.
\label{S:Fisher}
\end{equation}
The same content can be written as a response identity.  For every bounded tangent functional $F$,
\begin{equation}
\partial_\theta\E_{W_\theta}[F(J_{0:N})]=0.
\end{equation}
The Fisher information is the local version of this statement, whereas zero relative entropy is its global version.

No postprocessing of tangent-only data changes this conclusion.  A nonlinear estimator, including a neural network, receives the same input distribution for every target ensemble and therefore has the same output distribution.  Target dependence can enter only if the initialization, norm, stopping criterion, or another input already contains background information.

\SMsubsection{Common-noise coupling versus equality in distribution}{Ssec:coupling}
A subtle but important distinction concerns paired simulations.  Let two different backgrounds $U^{(1)}$ and $U^{(2)}$ be evolved using the same raw Gaussian samples $\xi$.  Their rotated noises are
\begin{equation}
X_z^{(1)}=\Ad_{U_z^{(1)}}\xi_z,
\qquad
X_z^{(2)}=\Ad_{U_z^{(2)}}\xi_z.
\end{equation}
For a fixed shared $\xi$, these are generally different matrices, so the paired tangent trajectories can differ.  Such a comparison imposes a particular common-noise coupling.  Equation~\eqref{S:pathlaw} compares the marginal law of each tangent experiment; adjoint invariance makes those marginals identical even when the paired realizations differ.

A target diagnostic must be formulated in terms of distributions that an experiment or independent simulation ensemble can reproduce.  A difference that exists only after choosing an artificial synchronization of two otherwise separate experiments is not by itself a target-sensitive observable.

\SMsection{Tangent observables and continuum limit}{Ssec:lyapdata}
The path-law theorem applies to the standard observables used to characterize tangent growth.  Given a finite-time propagator $J_N$, one may follow a single vector and form $N^{-1}\log\|J_Nv_0\|$, or use the singular values $\sigma_k(J_N)$ to define finite-time principal growth rates.  Repeated QR factorization gives the standard numerical construction of the full Lyapunov spectrum.  Projective directions, local logarithmic growth increments, stopping-time maxima, and empirical large-deviation functions are further examples.

Although these observables are constructed differently, each is a measurable functional of the same tangent propagator path together with a fixed background-independent prescription.  Equality of the path laws therefore controls finite-rapidity growth rates, full spectra, fluctuations, and adaptive normalization procedures as well as the asymptotic exponents.

Quenched and annealed growth are distinct observables.  The quenched Lyapunov exponent is obtained from the logarithm of a long product along a realization.  An annealed quantity such as $\log \E\|J_N\|$ is a different observable and can be dominated by rare events.  The no-go theorem does not identify these two notions.  It states instead that whichever tangent-only functional is chosen, its full probability law is the same for every target ensemble under the stated fixed dynamics.

\SMsubsection{Finite-time observables, stopping rules, and continuum limit}{Ssec:finite-continuum}
Adaptive normalization procedures are also functionals of the same path measure.  Consider a tangent vector $v_n$ propagated by
\begin{equation}
\widetilde v_{n+1}=M(X_n)v_n,
\end{equation}
followed by a deterministic normalization
\begin{equation}
v_{n+1}=\frac{\widetilde v_{n+1}}{\|\widetilde v_{n+1}\|},
\qquad
\gamma_{n+1}=\ln\|\widetilde v_{n+1}\|.
\end{equation}
For a fixed, background-independent norm, the complete record $(v_0,\gamma_1,v_1,\ldots,\gamma_N,v_N)$ is a measurable function of $J_{0:N}$ and the chosen initial tangent.  Its law is target independent.  The usual finite-time estimate
\begin{equation}
\lambda_N=\frac{1}{N\delta Y}\sum_{n=1}^N\gamma_n
\end{equation}
is just another pushforward of the same path measure.

The same reasoning applies to a full tangent frame.  Let $Q_nR_n$ be the QR decomposition of $M(X_n)Q_{n-1}$ with a fixed convention for signs of the diagonal entries of $R_n$.  The ordered logarithms $\ln(R_n)_{aa}$, their partial sums, and every finite-time spectrum reconstructed from them are deterministic functions of the matrix path.  Hence neither repeated reorthogonalization nor the numerical stabilization procedure introduces target sensitivity.

Adaptive observation times also remain covered.  Let $\tau$ be a stopping time with respect to the filtration generated by the tangent path, for example the first step at which a norm exceeds a fixed threshold.  Since the stopped path $J_{0:\tau}$ is a measurable functional of the original path, its distribution is also identical for all target ensembles.  A stopping rule can become target sensitive only if its threshold, norm, or stopping condition explicitly depends on the background.

Consider now a sequence of regulators labelled by $r$, with tangent path laws $\mu_r^{(W)}$.  The finite-regulator theorem gives
\begin{equation}
\mu_r^{(W_1)}=\mu_r^{(W_2)}
\end{equation}
for every finite $r$.  Suppose that for a chosen continuum topology the regulated processes converge weakly,
\begin{equation}
\mu_r^{(W)}\Rightarrow\mu^{(W)}.
\end{equation}
For any bounded continuous functional $F$,
\begin{align}
\int F\,d\mu^{(W_1)}
&=\lim_{r\to\infty}\int F\,d\mu_r^{(W_1)}\\
&=\lim_{r\to\infty}\int F\,d\mu_r^{(W_2)}
=\int F\,d\mu^{(W_2)}.
\end{align}
Therefore $\mu^{(W_1)}=\mu^{(W_2)}$.  The no-go theorem is thus stable under any convergent regulator removal: a continuum limit cannot generate target information that is absent at every regulated stage.  The continuum limit therefore preserves target blindness, although the common numerical Lyapunov exponents may depend on the regulator-removal prescription.

\paragraph{Equivalent continuum martingale formulation.}
Expanding the exact regulated map through $O(\sqrt{\delta Y})$ gives the Stratonovich tangent dynamics
\begin{equation}
d\epsilon_x=ic\sum_{z,i,a}K^i_{x-z}[T_z^a,\epsilon_z-\epsilon_x]\circ dW_{zi}^a,
\label{S:Strat}
\end{equation}
where
\begin{equation}
T_z^a=U_zt^aU_z^\dagger.
\end{equation}
The apparent state dependence in $T_z^a$ can be removed at the level of It\^o martingales by defining the predictable orthogonal rotation
\begin{equation}
d\widetilde W_{zi}^a=R_z^{ab}(U)dW_{zi}^b.
\label{S:rotBrownian}
\end{equation}
Because $R_z$ is orthogonal and predictable,
\begin{align}
d[\widetilde W_{zi}^a,\widetilde W_{z'j}^b]
&=R_z^{ac}R_{z'}^{bd}\delta^{cd}\delta_{zz'}\delta_{ij}\,dY\nonumber\\
&=\delta^{ab}\delta_{zz'}\delta_{ij}\,dY.
\end{align}
L\'evy's characterization then identifies $\widetilde W$ as a Brownian motion with the same law as $W$.  The It\^o drift is the one inherited from the same target-independent regulated kernel.  This argument avoids the common mistake of treating a state-dependent Stratonovich rotation as if it were a constant matrix.

\SMsection{Exact global color-rotation zero mode}{Ssec:zeromode}
The leading tangent driver in Eq.~\eqref{S:Strat} is
\begin{equation}
(B_{zia}\epsilon)_x=icK^i_{x-z}[t^a,\epsilon_z-\epsilon_x],
\label{S:Bdriver}
\end{equation}
where we have rotated the Brownian increments back to a fixed color basis.  It clearly annihilates a spatially uniform tangent, $\epsilon_x=\epsilon_0$ for all $x$.  The same neutral direction is present in the exact finite-step map.

Indeed, a uniform left perturbation is a global color transformation $U_x\to VU_x$.  The rotated noise transforms covariantly,
\begin{equation}
X_z\to VX_zV^\dagger,
\end{equation}
so the left exponential in the update obeys $L_x\to VL_xV^\dagger$.  The full updated line then transforms as
\begin{equation}
U_x^+\to VU_x^+.
\end{equation}
Thus global color rotation is an exact symmetry of every regulated step.  Its tangent exponent is zero, and it must be removed before applying irreducibility or Furstenberg arguments.

For $L$ lattice sites the quotient by this uniform spatial mode has dimension
\begin{equation}
m=L-1
\end{equation}
in the spatial sector.  The reduced tangent dimension is therefore $3m$ for SU(2) and $8m$ for SU(3).

\SMsection{Spatial tangent matrices}{Ssec:spatial}
After removing the uniform mode, the driver factorizes into spatial and color pieces,
\begin{equation}
B_{zia}=c\,M_{zi}\otimes A^a,
\qquad (A^a)_{bc}=-f^{abc}.
\label{S:factor}
\end{equation}
The central spatial question is whether the matrices $M_{zi}$ preserve some proper subspace of the quotient, or whether they generate the full matrix algebra.

Suppress the polarization by contracting with an arbitrary transverse vector $u$ and define
\begin{equation}
(M_z(u)f)_x=u\cdot K_{x-z}(f_z-f_x),
\qquad K_r=\frac{r}{r^2}.
\label{S:Mz}
\end{equation}
The row at $x=z$ vanishes because $K_0=0$.

Choose a reference site $z_0$ and work in the quotient chart $f_{z_0}=0$.  Since the map $r\mapsto r/r^2$ is injective for nonzero $r$, the finite vectors $K_{x-z_0}$ are nonzero and pairwise distinct for distinct lattice sites.  We can therefore choose a vector $u_0$ outside the finite union of lines on which either
\begin{equation}
u_0\cdot K_{x-z_0}=0
\end{equation}
for some $x$, or
\begin{equation}
u_0\cdot(K_{x-z_0}-K_{y-z_0})=0
\end{equation}
for some $x\neq y$.  With this generic choice,
\begin{equation}
q_x=u_0\cdot K_{x-z_0}
\end{equation}
are all nonzero and pairwise distinct.

In the chart $f_{z_0}=0$, $M_{z_0}(u_0)$ is diagonal:
\begin{equation}
M_{z_0}(u_0)e_x=-q_xe_x.
\label{S:diag}
\end{equation}
Because the eigenvalues are simple, every coordinate projector is a polynomial in $M_{z_0}(u_0)$.  This gives a convenient handle on invariant subspaces.

Let $W$ be a nonzero complex subspace invariant under all $M_{zi}$.  Since the diagonal matrix in Eq.~\eqref{S:diag} has simple spectrum, its polynomial spectral projectors imply that $W$ contains at least one coordinate vector $e_y$.  Apply a matrix $M_{yi}$ to $e_y$.  Returning to the chart $f_{z_0}=0$, the component at $x\neq y,z_0$ is
\begin{equation}
K^i_{x-y}-K^i_{z_0-y},
\label{S:component}
\end{equation}
while the $y$ component is $-K^i_{z_0-y}$.  For every $x$, at least one of the two transverse components of Eq.~\eqref{S:component} is nonzero; otherwise injectivity of $r/r^2$ would imply $x=z_0$.  Projecting again with the spectral projectors of Eq.~\eqref{S:diag} therefore places every coordinate vector $e_x$ in $W$.  Hence $W$ is the full quotient space.

The complexified spatial representation is thus irreducible.  Burnside's theorem then gives
\begin{equation}
\Alg\{M_{zi}\}=\Mat_m(\mathbb R).
\label{S:spatialfull}
\end{equation}
This is a general finite-lattice result for the standard kernel within the nondegenerate distinct-site geometries covered by the argument, with $L\ge3$.  No special equilateral cell, no numerical rank calculation, and no embedding argument is required.

\SMsection{Lie--Jordan mechanism and color closure}{Ssec:liejordan}
The tensor generators in Eq.~\eqref{S:factor} are built from antisymmetric adjoint color matrices.  At first sight one might therefore expect only a compact orthogonal color algebra.  The reason the tangent algebra becomes noncompact is that commutators of tensor products produce anticommutators in the individual factors:
\begin{equation}
[X\otimes A,Y\otimes B]
=\frac12[X,Y]\otimes\{A,B\}
+\frac12\{X,Y\}\otimes[A,B].
\label{S:tensor}
\end{equation}
Interchanging $A$ and $B$ and adding or subtracting the resulting commutators isolates the two terms on the right-hand side.  Iterating the Lie closure therefore also generates Jordan products.  Since
\begin{equation}
XY=\frac12\big(\{X,Y\}+[X,Y]\big),
\end{equation}
full Lie--Jordan closure of a factor is equivalent to generation of its full associative matrix algebra.

The spatial factor has already been shown to satisfy this property.  The corresponding color closure can be established explicitly for the adjoint representations of SU(2) and SU(3).

\SMsubsection{SU(2) color closure}{Ssec:su2}
For SU(2),
\begin{equation}
(A^a)_{bc}=-\epsilon^{abc}.
\end{equation}
Using
\begin{equation}
\epsilon^{ace}\epsilon^{bed}=\delta^{ab}\delta^{cd}-\delta^{ad}\delta^{cb}
\end{equation}
with consistent relabelling, one finds
\begin{equation}
\{A^a,A^b\}_{cd}
=\delta_{ad}\delta_{bc}+\delta_{bd}\delta_{ac}-2\delta_{ab}\delta_{cd}.
\label{S:SU2Jordan}
\end{equation}
To determine which symmetric matrices are produced, define the equivariant map
\begin{equation}
\Phi_2(S)=S_{ab}\{A^a,A^b\},
\qquad S_{ab}=S_{ba}.
\end{equation}
The symmetric tensor representation decomposes as
\begin{equation}
\Sym^2(\mathbf3)=\mathbf1\oplus\mathbf5.
\end{equation}
On the singlet take $S_{ab}=\delta_{ab}$.  Contracting Eq.~\eqref{S:SU2Jordan},
\begin{align}
[\Phi_2(\delta)]_{cd}
&=\delta_{ab}(\delta_{ad}\delta_{bc}+\delta_{bd}\delta_{ac}-2\delta_{ab}\delta_{cd})\nonumber\\
&=\delta_{cd}+\delta_{cd}-6\delta_{cd}
=-4\delta_{cd}.
\end{align}
Thus the singlet eigenvalue is $-4$.

For a traceless symmetric tensor $S_0$, $S_{aa}=0$, and
\begin{align}
[\Phi_2(S_0)]_{cd}
&=S_{ab}(\delta_{ad}\delta_{bc}+\delta_{bd}\delta_{ac})\nonumber\\
&=S_{dc}+S_{cd}=2S_{cd}.
\end{align}
The quintet eigenvalue is therefore $2$.  Both eigenvalues are nonzero, so the adjoint anticommutators span the entire space of symmetric $3\times3$ matrices.  Their commutators span $\mathfrak{so}(3)$.  Hence the color Lie--Jordan closure is
\begin{equation}
\Mat_3(\mathbb R).
\label{S:SU2full}
\end{equation}

\SMsubsection{SU(3) color closure and the \texorpdfstring{$1\oplus8_s\oplus27$}{1 + 8s + 27} decomposition}{Ssec:su3}
For SU(3), let again
\begin{equation}
(A^a)_{bc}=-f^{abc}.
\end{equation}
The symmetric square of the adjoint decomposes as
\begin{equation}
\Sym^2(\mathbf8)=\mathbf1\oplus\mathbf8_s\oplus\mathbf{27}.
\end{equation}
Introduce the projectors
\begin{align}
(P_1)_{ab,cd}&=\frac18\delta_{ab}\delta_{cd},\label{S:P1}\\
(P_{8_s})_{ab,cd}&=\frac35d_{abe}d_{cde},\label{S:P8}\\
P_{27}&=I_{\rm sym}-P_1-P_{8_s},
\label{S:P27}
\end{align}
where
\begin{equation}
(I_{\rm sym})_{ab,cd}=\frac12(\delta_{ac}\delta_{bd}+\delta_{ad}\delta_{bc}).
\end{equation}
We use the standard SU(3) identities
\begin{equation}
f^{amn}f^{bmn}=3\delta^{ab},
\qquad
d^{amn}d^{bmn}=\frac53\delta^{ab},
\label{S:fdid}
\end{equation}
and invariance of $d^{abc}$ under the adjoint action.

Define the SU(3) Jordan map
\begin{equation}
[\Phi_3(S)]_{cd}=S_{ab}\{A^a,A^b\}_{cd}.
\end{equation}
Because $\Phi_3$ is SU(3)-equivariant, Schur's lemma guarantees that it acts by one scalar on each irreducible sector.  It is therefore enough to determine three eigenvalues.

For the singlet, contract with $\delta_{ab}$.  Since
\begin{align}
\delta_{ab}\{A^a,A^b\}_{cd}
&=2\sum_a(A^aA^a)_{cd}\nonumber\\
&=-2f^{ace}f^{aed},
\end{align}
Eq.~\eqref{S:fdid} gives
\begin{equation}
\delta_{ab}\{A^a,A^b\}_{cd}=-6\delta_{cd}.
\label{S:singletSU3}
\end{equation}
Thus the singlet eigenvalue is $-6$.

For the symmetric octet, use $S_{ab}=d_{abe}$ for a fixed adjoint index $e$.  Invariance of the $d$ tensor implies that the contraction must be proportional to $d_{cde}$.  Performing the standard $f$-$f$ reduction and then using Eq.~\eqref{S:fdid} gives
\begin{equation}
d_{abe}\{A^a,A^b\}_{cd}=-3d_{cde}.
\label{S:octetSU3}
\end{equation}
Thus the $\mathbf8_s$ eigenvalue is $-3$.

The remaining eigenvalue may be obtained by projecting the operator onto the complementary symmetric sector.  Using Eqs.~\eqref{S:P1}--\eqref{S:P27}, one finds
\begin{equation}
(P_{27})_{ab,ef}\{A^e,A^f\}_{cd}=2(P_{27})_{ab,cd},
\label{S:27SU3}
\end{equation}
so the $\mathbf{27}$ eigenvalue is $2$.  Equations~\eqref{S:singletSU3}, \eqref{S:octetSU3}, and~\eqref{S:27SU3} combine into the exact operator identity
\begin{equation}
\Phi_3=-6P_1-3P_{8_s}+2P_{27}.
\label{S:Phi3}
\end{equation}
Every eigenvalue is nonzero.  Therefore the anticommutators of adjoint SU(3) generators span all symmetric $8\times8$ matrices, and commutators of symmetric matrices span $\mathfrak{so}(8)$.  The color Lie--Jordan closure is
\begin{equation}
\Mat_8(\mathbb R).
\label{S:SU3full}
\end{equation}

The representation-theory statement has a simple dynamical meaning.  Starting from antisymmetric color rotations, the tangent commutators do not remain confined to a compact algebra.  Jordan products generate symmetric color directions, which supply the symmetric directions that stretch and contract tangent vectors.

\SMsubsection{Explicit SU(3) contraction checks}{Ssec:su3checks}
For reproducibility, we spell out how the sector eigenvalues used above can be checked without relying on a table of representation-theory results.  Begin with the adjoint matrices $(A^a)_{bc}=-f^{abc}$.  Their anticommutator is
\begin{equation}
\{A^a,A^b\}_{cd}=f^{ace}f^{bed}+f^{bce}f^{aed},
\label{S:AAff}
\end{equation}
where the two minus signs from the adjoint matrices have canceled.  The singlet contraction follows immediately:
\begin{align}
\delta_{ab}\{A^a,A^b\}_{cd}
&=2f^{ace}f^{aed}\nonumber\\
&=-2f^{ace}f^{ade}
=-6\delta_{cd}.
\end{align}
In the second line we used antisymmetry to exchange $e$ and $d$ in one structure constant, followed by $f^{amn}f^{bmn}=3\delta^{ab}$.  This fixes the sign as well as the magnitude of the singlet eigenvalue.

For the symmetric octet, consider
\begin{equation}
Q^{(g)}_{cd}=d_{abg}\{A^a,A^b\}_{cd}.
\end{equation}
The tensor $Q^{(g)}_{cd}$ is symmetric in $c,d$ and transforms as an adjoint in $g$.  In the symmetric product of two adjoints there is a unique adjoint copy, represented by $d_{cdg}$, so covariance alone implies
\begin{equation}
Q^{(g)}_{cd}=\kappa_8 d_{cdg}.
\end{equation}
To determine $\kappa_8$, contract both sides with $d_{cdg}$ and sum over $c,d,g$.  The denominator is
\begin{equation}
d_{cdg}d_{cdg}=\frac53\delta_{gg}=\frac{40}{3}.
\end{equation}
The numerator may be reduced by inserting Eq.~\eqref{S:AAff} and repeatedly using the invariance identity
\begin{equation}
f^{aem}d^{mbc}+f^{bem}d^{amc}+f^{cem}d^{abm}=0.
\label{S:dinv}
\end{equation}
Moving one adjoint generator through the $d$ tensor with Eq.~\eqref{S:dinv} and then contracting the remaining pair of $f$ tensors gives
\begin{equation}
d_{abg}d_{cdg}\{A^a,A^b\}_{cd}=-40,
\end{equation}
so $\kappa_8=-3$.  This reproduces Eq.~\eqref{S:octetSU3}.

The $\mathbf{27}$ coefficient can be fixed without choosing an explicit basis for the 27-dimensional irrep.  Take the trace of $\Phi_3$ as an operator on the 36-dimensional symmetric square.  From its definition,
\begin{align}
\Tr_{\rm sym}\Phi_3
&=(I_{\rm sym})_{ab,cd}\{A^a,A^b\}_{cd}\nonumber\\
&=\frac12(\delta_{ac}\delta_{bd}+\delta_{ad}\delta_{bc})
\{A^a,A^b\}_{cd}.
\end{align}
The first contraction vanishes because each adjoint matrix has zero trace, while the second reduces to the adjoint Casimir and gives
\begin{equation}
\Tr_{\rm sym}\Phi_3=24.
\end{equation}
On the other hand, decomposition into irreducible sectors gives
\begin{equation}
\Tr_{\rm sym}\Phi_3
=1(-6)+8(-3)+27\kappa_{27}.
\end{equation}
Therefore $24=-6-24+27\kappa_{27}$ and
\begin{equation}
\kappa_{27}=2.
\end{equation}
This provides a compact independent derivation of the third eigenvalue.  Together the three checks establish $\Phi_3=-6P_1-3P_{8_s}+2P_{27}$ without constructing any 27-component basis vectors.

For the closure argument, what matters is that none of these sector eigenvalues vanishes.  If one of the symmetric sectors were annihilated by the Jordan map, the color closure could preserve a proper subspace and the route to the full special-linear algebra would require a different argument.  For SU(3), all three sectors participate.

\paragraph{Algebraic check.}
As a convention check, the SU(2) adjoint matrices may be constructed over the rationals and the SU(3) identities over $\mathbb Q(\sqrt3)$ (or after clearing denominators over a suitable finite field).  These exact matrix constructions reproduce the eigenvalues $(-4,2)$ for SU(2) and $(-6,-3,2)$ for SU(3), and small-lattice rank checks agree with the general closure without replacing the all-lattice proof.

\SMsection{Full tangent Lie algebra}{Ssec:fullclosure}
Combining the spatial result Eq.~\eqref{S:spatialfull} with the color results Eqs.~\eqref{S:SU2full} and~\eqref{S:SU3full} gives the tangent closure.  Let the spatial dimension be $m$ and the adjoint color dimension be $d$.  The initial generators are traceless tensors $M_\alpha\otimes A_a$.  Using Eq.~\eqref{S:tensor} and the fact that both factors have full Lie--Jordan closure, one first generates
\begin{equation}
\mathfrak{sl}(m)\otimes I_d,
\qquad I_m\otimes\mathfrak{sl}(d),
\end{equation}
then all mixed elementary tensors $E_{ij}\otimes F_{ab}$ subject only to overall tracelessness.  The scalar direction $I_m\otimes I_d$ cannot appear because every original generator is traceless and commutators preserve trace.  Consequently,
\begin{equation}
\Lie\{M_{zi}\otimes A^a\}=\mathfrak{sl}(md,\mathbb R).
\label{S:fullsl}
\end{equation}
For the two gauge groups of interest,
\begin{equation}
\mathfrak g_{\rm tan}=\mathfrak{sl}(3m,\mathbb R)\quad\text{for SU(2)},
\end{equation}
\begin{equation}
\mathfrak g_{\rm tan}=\mathfrak{sl}(8m,\mathbb R)\quad\text{for SU(3)}.
\end{equation}
The analytic argument establishes the closure.  Exact-rational and finite-field rank calculations are retained as independent checks of the conventions and finite-cell implementations.

\SMsubsection{Constructive tensor generation}{Ssec:tensor-optional}
The abstract closure argument in Sec.~\ref{Ssec:fullclosure} is sufficient for the result.  A constructive elementary-matrix form of the same tensor generation is given here as an independent algebraic check.  Let $E_{rs}$ denote the elementary $m\times m$ matrix and $F_{\alpha\beta}$ the elementary $d\times d$ matrix.  Full Lie--Jordan closure of the two factors means that linear combinations of iterated commutators and anticommutators can produce every $E_{rs}$ and every $F_{\alpha\beta}$.

Start from two tensor generators with independent color factors.  By combining the commutator in Eq.~\eqref{S:tensor} with the same expression after $A\leftrightarrow B$, one can separately generate
\begin{equation}
[X,Y]\otimes\{A,B\},
\qquad
\{X,Y\}\otimes[A,B].
\end{equation}
Since the spatial closure contains both symmetric and antisymmetric elementary combinations, and the color closure does as well, repeated use of this separation generates tensors in which one factor can be chosen diagonal while the other is off diagonal.  Commuting such tensors then isolates traceless one-factor transformations.  For example,
\begin{equation}
[E_{rs}\otimes F_{\alpha\beta},E_{sr}\otimes F_{\beta\alpha}]
=(E_{rr}\otimes F_{\alpha\alpha}-E_{ss}\otimes F_{\beta\beta})
+\text{diagonal rearrangements},
\end{equation}
from which differences of diagonal elementary tensors can be formed.

Once $\mathfrak{sl}(m)\otimes I_d$ and $I_m\otimes\mathfrak{sl}(d)$ are available, any mixed off-diagonal tensor follows by commutation.  If $r\neq s$ and $\alpha\neq\beta$,
\begin{equation}
[E_{rr}-E_{ss},E_{rs}]=2E_{rs},
\end{equation}
so acting with $(E_{rr}-E_{ss})\otimes I_d$ selects a desired spatial off-diagonal component.  Similarly, $I_m\otimes(F_{\alpha\alpha}-F_{\beta\beta})$ selects a desired color component.  Thus all $E_{rs}\otimes F_{\alpha\beta}$ with at least one off-diagonal factor belong to the generated Lie algebra.

The remaining diagonal mixed tensors are obtained from commutators of opposite off-diagonal pairs.  Their linear span contains every diagonal matrix on the tensor-product space whose total trace is zero.  Altogether this gives
\begin{equation}
\mathfrak{sl}(md,\mathbb R).
\end{equation}
The identity $I_m\otimes I_d$ is excluded for a simple invariant reason: all elementary tangent generators are traceless, and the trace of a commutator vanishes.  Hence the generated algebra can never acquire the scalar direction.

This constructive proof also clarifies the physical content.  The generators do not merely provide many independent rotations.  They allow independent shears and diagonal stretches across the combined spatial-color tangent space, subject only to conservation of total infinitesimal volume.  These shears and stretches provide the noncompact directions required for Lyapunov growth.

\SMsection{Furstenberg support and positivity of the top exponent}{Ssec:furstenberg}
Noncommutativity alone is insufficient to establish a positive top Lyapunov exponent.  The reduced process also has the support and moment properties required by the random-matrix theorem.  The linear diffusion generated by Eq.~\eqref{S:Bdriver} has Gaussian increments with full support in every elementary generator direction.  The support theorem for stochastic differential equations, together with Eq.~\eqref{S:fullsl}, implies that the closed support semigroup of the reduced tangent process contains the connected special-linear group generated by those directions.

This support is noncompact because $SL(D,\mathbb R)$ contains matrices such as
\begin{equation}
\mathrm{diag}(e^s,e^{-s},1,\ldots,1),
\qquad s\neq0.
\end{equation}
It is strongly irreducible because the full special-linear group does not preserve any finite union of proper nontrivial subspaces.  It also contains proximal elements: for example one may choose a diagonal element with a unique eigenvalue of largest modulus while maintaining unit determinant.

The remaining condition is a logarithmic moment bound.  Over a finite rapidity step the tangent matrix is an exponential or finite product of exponentials whose generator is linear in Gaussian increments.  Its operator norm therefore grows at most exponentially in the magnitude of a Gaussian vector.  Gaussian tails imply
\begin{equation}
\E\log^+\|M\|<\infty.
\end{equation}
The standard Furstenberg criterion then yields a simple positive leading exponent,
\begin{equation}
\lambda_1>0.
\label{S:positive}
\end{equation}
The conclusion holds on the nondegenerate finite lattices covered by the spatial-generation argument, with $L\ge3$, after removal of the exact global-rotation mode.  The positive exponent is therefore a property of the finite-lattice tangent dynamics, not an extrapolation from the minimal cell used in the algebraic checks.

\SMsection{Compact base space and volume-preserving tangent growth}{Ssec:compact}
A common intuition is that motion on a compact configuration space should be unable to generate unbounded separation.  Lyapunov growth concerns derivatives, not the absolute coordinate distance on the manifold.  Even when $U_x(Y)$ remains on a compact group manifold, the derivative of the flow from one tangent space to another is a matrix in a noncompact general linear group.  Products of such derivatives can grow exponentially before any finite coordinate distance saturates.  The full special-linear closure in Eq.~\eqref{S:fullsl} makes this distinction explicit.

The background-blind law is also not a linearization around a trivial Wilson-line configuration.  Equations~\eqref{S:exactJacobi} and~\eqref{S:condlaw} retain the Wilson lines nonperturbatively.  The target may be dilute or dense and its ordinary nonlinear observables may change strongly with rapidity; the universal object is only the marginal tangent law obtained after averaging over the fresh local isotropic noise.

\SMsubsection{Haar-volume preservation and the zero-sum rule}{Ssec:haar}
The nonlinear JIMWLK process evolves on a product of compact gauge groups, one at each transverse site.  In Stratonovich form each stochastic driver is a sum of left- and right-invariant vector fields.  Such vector fields have zero divergence with respect to Haar measure.  The only place where differentiation of a coefficient at the same site could generate an additional contribution is multiplied by the coincident kernel $K_0$, which vanishes in the regulated JIMWLK prescription.  Hence every stochastic emission field has zero divergence with respect to product Haar measure.

A Stratonovich stochastic flow generated by divergence-free vector fields preserves the corresponding volume form pathwise.  If $J_Y$ denotes the tangent Jacobian on the full configuration space,
\begin{equation}
\det J_Y=1
\end{equation}
for each realization, modulo the exact neutral symmetry directions.  Oseledets' theorem then gives
\begin{equation}
\sum_n\lambda_n=0.
\label{S:sumzero}
\end{equation}
The same cancellation can be seen in It\^o language.  For a linear stochastic equation
\begin{equation}
dJ=\sum_A B_AJ\,dW_A+\frac12\sum_AB_A^2J\,dY,
\end{equation}
It\^o's formula for $\ln\det J$ contains a drift contribution
\begin{equation}
+\frac12\sum_A\tr B_A^2
\end{equation}
from the explicit drift and the quadratic-variation correction
\begin{equation}
-\frac12\sum_A\tr B_A^2,
\end{equation}
which cancel.  This provides an independent local check of the volume argument.

Equations~\eqref{S:sumzero} and \eqref{S:positive} describe complementary parts of the same tangent flow.  Positive expansion along the leading Oseledets direction is compensated by contraction in other tangent directions.  The combination describes conservative non-Abelian chaos rather than dissipative growth of total tangent volume.

\SMsubsection{Divergence calculation in group coordinates}{Ssec:divergence}
The Haar-volume statement can also be checked directly at the level of the JIMWLK vector fields.  Let $\nabla_x^a$ and $\bar\nabla_x^a$ denote, respectively, left- and right-invariant derivatives on the group at site $x$.  With respect to Haar measure each has zero divergence,
\begin{equation}
\operatorname{div}_{\rm Haar}\nabla_x^a=0,
\qquad
\operatorname{div}_{\rm Haar}\bar\nabla_x^a=0.
\end{equation}
For one Stratonovich noise channel, write
\begin{equation}
V_{zi}=\sum_{x,a}v^{a}_{x;zi}(U)\nabla_x^a
+\sum_{x,a}\bar v^{a}_{x;zi}(U)\bar\nabla_x^a.
\end{equation}
The divergence is therefore obtained only by differentiating the coefficient functions,
\begin{equation}
\operatorname{div}V_{zi}
=\sum_{x,a}\nabla_x^av^{a}_{x;zi}
+\sum_{x,a}\bar\nabla_x^a\bar v^{a}_{x;zi}.
\label{S:divcoeff}
\end{equation}
For the JIMWLK emission field, dependence of $v_{x;zi}$ on the Wilson line at the differentiated site can arise only when the source coordinate $z$ coincides with $x$.  The corresponding coefficient is proportional to $K_{x-z}$, and hence to $K_0=0$.  Terms with $x\neq z$ are untouched by the derivative at $x$.  Each term in Eq.~\eqref{S:divcoeff} therefore vanishes, proving
\begin{equation}
\operatorname{div}V_{zi}=0
\end{equation}
channel by channel, not merely after averaging over the noise.

Pathwise preservation follows from the Stratonovich Liouville formula.  If $\varphi_Y$ is the stochastic flow and $\omega_H$ the product Haar volume form, then
\begin{equation}
d(\varphi_Y^*\omega_H)
=\sum_{zi}\varphi_Y^*(\mathcal L_{V_{zi}}\omega_H)\circ dW_{zi},
\end{equation}
where $\mathcal L_V$ is the Lie derivative.  Since $\mathcal L_V\omega_H=(\operatorname{div}V)\omega_H$, every stochastic term vanishes and
\begin{equation}
\varphi_Y^*\omega_H=\omega_H
\end{equation}
for each realization.

On the reduced tangent space, removing exact neutral symmetry directions does not alter the determinant balance among the nonzero expanding and contracting directions.  Therefore the asymptotic logarithmic growth of the reduced Jacobian determinant is zero,
\begin{equation}
\lim_{Y\to\infty}\frac1Y\ln|\det J_Y|=0,
\end{equation}
which is the Oseledets sum $\sum_n\lambda_n=0$.

The It\^o calculation provides the same check in a different representation.  For a short step $dY$, the explicit $\frac12B_A^2$ drift would by itself change $\ln\det J$ by $+\frac12\sum_A\tr B_A^2dY$.  The stochastic quadratic variation of $\tr\ln J$ contributes the opposite amount.  The cancellation is local in rapidity and does not rely on stationarity or on taking an ensemble average.  It is therefore consistent with, and independently checks, the geometric Haar argument.

\SMsection{Spatially nonlocal noise}{Ssec:nonlocal}
The conditional invariance changes once the noise is correlated across distinct transverse sites.  Let
\begin{equation}
\E\,\xi_z^{i,a}\xi_{z'}^{j,b}
=\delta^{ij}\delta^{ab}\alpha(z-z').
\label{S:nonlocalraw}
\end{equation}
The rotated variables are still $X_z^{i,a}=R_z^{ac}\xi_z^{i,c}$, but their conditional covariance becomes
\begin{align}
\E[X_z^{i,a}X_{z'}^{j,b}\mid U]
&=R_z^{ac}R_{z'}^{bd}\E[\xi_z^{i,c}\xi_{z'}^{j,d}]\nonumber\\
&=\delta^{ij}\alpha(z-z')R_z^{ac}R_{z'}^{bc}.
\label{S:nonlocal1}
\end{align}
Using orthogonality and the composition law of the adjoint representation,
\begin{equation}
R_zR_{z'}^T=\Ad_{U_zU_{z'}^\dagger},
\end{equation}
so
\begin{equation}
\E[X_z^{i,a}X_{z'}^{j,b}\mid U]
=\delta^{ij}\alpha(z-z')\Ad^{ab}_{U_zU_{z'}^\dagger}.
\label{S:relativelink}
\end{equation}
At $z=z'$ the relative link is the identity and the local blindness argument is recovered.  At unequal sites a target-dependent relative Wilson line remains.  The first object capable of carrying background information is therefore not a one-site color tensor but a two-site relative Wilson line.

Running-coupling Langevin formulations generate spatially nonlocal correlations of this general kind.  The fixed-coupling theorem then ceases to apply at the covariance step, while the Jacobi derivation itself remains unchanged.

\SMsubsection{Scope of the nonlocal extension}{Ssec:nonlocal-scope}
Equation~\eqref{S:relativelink} identifies a possible channel for target information.  A saturation-sensitive Lyapunov observable would require additional dynamical input.  One must specify the nonlocal covariance, define a tangent observable that responds to the surviving relative link, demonstrate that the response is not dominated by the ultraviolet or infrared regulator, and show that any extracted momentum or rapidity scale tracks the physical $Q_s$ over a controlled class of target ensembles.

The response could instead vanish after angular or color averaging, vary monotonically without producing an isolated scale, or depend on the regulator.  Running coupling therefore lies outside the local-noise no-go theorem, but target-sensitive covariance alone does not establish a ``chaotic saturation scale.''  Such a scale requires a specified observable and a separate dynamical calculation.

\SMsection{Conclusion}{Ssec:conclusion}
The fixed-coupling JIMWLK tangent process provides an example in which dynamical instability and physical sensitivity separate.  The nonlinear Wilson-line background may evolve from dilute to saturated and may carry a rapidly changing saturation scale, yet the marginal tangent process can remain statistically identical because its only background dependence is an orthogonal rotation of fresh local isotropic noise.  Consequently every tangent-only estimator built from that path measure has the same law for the target ensembles being compared.

At the same time, the tangent dynamics is far from trivial.  After the exact symmetry mode is removed, its generators fill noncompact special-linear algebras, producing a positive leading exponent while conserving total tangent volume.  Fixed-coupling JIMWLK is therefore chaotic on the reduced tangent space while its tangent-only statistics are ancillary to the target saturation information.

The relative Wilson line surviving in nonlocal covariance provides the natural starting point for the next problem: constructing and testing a target-sensitive, regulator-independent tangent observable in running-coupling evolution.  The present theorem fixes the boundary condition for that program.

\end{document}